\documentclass[final]{iopart}
\usepackage{graphicx,subfigure,iopams}
\begin{document}

\title[Effect of strong correlations on the high energy anomaly]{Effect of strong correlations on the high energy anomaly 
in hole- and electron-doped high-$T_{c}$ superconductors}

\author{B. Moritz$^{1}$, F. Schmitt$^{2,3}$, W. Meevasana$^{2,4}$, S. Johnston$^{1,5}$, E. M. Motoyama$^{2,4}$, M. Greven$^{1,2,3}$, D. H. Lu$^{6}$,
C. Kim$^{7}$, R. T. Scalettar$^{8}$, Z.-X. Shen$^{1,2,3,4}$, T. P. Devereaux$^{1}$}

\address{$^{1}$ Stanford Institute for Materials and Energy Science, SLAC National Accelerator Laboratory and Stanford University, Stanford, CA 94305, USA}
\address{$^{2}$ Geballe Laboratory for Advanced Materials, Stanford University, Stanford, CA 94305, USA}
\address{$^{3}$ Department of Applied Physics, Stanford University, Stanford, CA 94305, USA}
\address{$^{4}$ Department of Physics, Stanford University, Stanford, CA 94305, USA}
\address{$^{5}$ Department of Physics and Astronomy, University of Waterloo, Waterloo, ON N2L 3G1, Canada}
\address{$^{6}$ Stanford Synchrotron Radiation Lightsource, SLAC National Accelerator Laboratory, Menlo Park, CA 94025, USA}
\address{$^{7}$ Institute of Physics and Applied Physics, Yonsei University, Seoul 120-749, Korea}
\address{$^{8}$ Physics Department, University of California - Davis, Davis, CA 95616, USA}

\date{\today}

\begin{abstract}
Recently, angle-resolved photoemission spectroscopy (ARPES) has been used to highlight an anomalously large band renormalization at high binding 
energies in cuprate superconductors: the high energy ``waterfall" or high energy anomaly (HEA).  This paper demonstrates, using a combination of new 
ARPES measurements and quantum Monte Carlo simulations, that the HEA is not simply the byproduct of matrix element effects, but rather represents a 
cross-over from a quasi-particle band at low binding energies near the Fermi level to valence bands at higher binding energy, assumed to be of strong 
oxygen character, in both hole- and electron-doped cuprates.  While photoemission matrix elements clearly play a role in changing the aesthetic 
appearance of the band dispersion, {\em i.e.} the ``waterfall"-like behavior, they provide an inadequate description for the physics that underlies 
the strong band renormalization giving rise to the HEA.  Model calculations of the single-band Hubbard Hamiltonian showcase the role played by 
correlations in the formation of the HEA and uncover significant differences in the HEA energy scale for hole- and electron-doped cuprates.  In 
addition, this approach properly captures the transfer of spectral weight accompanying both hole and electron doping in a correlated material and 
provides a unifying description of the HEA across both sides of the cuprate phase diagram.
\end{abstract}



\maketitle

\section{Introduction}
Over the last decade significant advancements have improved the momentum and energy resolution of angle-resolved photoemission spectroscopy (ARPES) 
\cite{ARPES_Rev} that provides access to the single-particle spectral function $A(\mathbf{k},\omega)$. While a significant probe of electronic 
structure in general, such advancements in ARPES have had a profound impact on the study of strongly correlated materials. A series of high 
resolution ARPES experiments at binding energies up to $\sim 1-1.5$ eV, made possible by these advancements, has revealed the presence of a high 
energy anomaly (HEA) in the dispersion of both hole- 
\cite{Graf_anomaly,Non_anomaly,Feng_anomaly,Valla_anomaly,Chang_anomaly,Borisenko_anomaly,Zhang_anomaly} and electron-doped  
(Sm$_{2-x}$Ce$_x$CuO$_{4}$ \cite{CKim_anomaly},  Nd$_{2-x}$Ce$_x$CuO$_{4}$ \cite{Fujimori_anomaly}, Pr$_{1-x}$LaCe$_{x}$CuO$_{4}$ \cite{Pan_anomaly}) 
high-$T_{c}$ superconductors.  In hole-doped compounds the anomaly,  at an energy $\sim 300$ meV, is present in different materials, doping values, 
and phases and more recently has been resolved in electron-doped compounds at approximately twice the energy scale.  A link for the HEA between hole- 
and electron-doped cuprates can be found in the similar ``waterfall" feature observed in the half-filled parent insulators \cite{Ronning_anomaly}. A 
number of theories have been advanced to explain the presence of this anomaly including spin-charge separation \cite{Graf_anomaly}, spin polarons 
\cite{Manousakis_HEA}, in-gap band-tails \cite{Alexandrov_HEA}, coupling to phonons \cite{Feng_anomaly}, plasmons \cite{Bansil_plasmon}, or 
paramagnons \cite{Valla_anomaly,QMC,Paramagnon_anomaly,Bansil_ME_effect}, photoemission matrix elements \cite{Borisenko_anomaly}, and strong 
correlation or ``Mott" physics \cite{Non_anomaly,DMFT,Tan_HEA,Lanczos,LDA+DMFT}.
These theories include
assertions that no HEA should be \cite{Lanczos} or is \cite{Kaminski_anomaly} present in electron-doped cuprates contrary to 
other experimental evidence \cite{CKim_anomaly,Fujimori_anomaly,Pan_anomaly}.  

The appearance of the HEA as a ``waterfall"- or kink-like feature provides conceptual appeal for certain theoretical explanations 
\cite{Feng_anomaly,Valla_anomaly,Bansil_plasmon,QMC,Paramagnon_anomaly,Bansil_ME_effect}.  However, such scenarios based on weak coupling to high 
energy bosonic modes have severe problems.  Attributing the HEA to electron-phonon coupling, similar to that used to describe the ``low-energy kink" 
in the cuprates \cite{Allesandra_ARPES,Devereaux_kink,ARPES_Zhou}, appears to be unlikely given the relatively large energy scale 
and its apparent lack of doping dependence.  
Coupling to paramagnons or spin fluctuations generally satisfies the energy for the HEA, based on the superexchange value 
J $\sim 100\ \mathrm{meV}$ in cuprates; however, such coupling fails to properly address the apparent 
damping of paramagnon modes with doping 
\cite{Sugai} and cannot account for the dichotomy in energy scales between hole- and electron-doped materials.  

In addition, the kink-like appearance gives way to ``band break-up" at low photon energy \cite{Zhang_anomaly} or a shallow band dispersion with a 
characteristic ``Y" appearance near the zone center in the second Brillouin zone (BZ) \cite{Borisenko_anomaly}, both effects primarily due to changes 
in photoemission matrix elements.  Photoemission matrix elements complicate the analysis of individual spectra through extrinsic effects 
\cite{Borisenko_anomaly}, 
but there remain intrinsic band renormalization effects creating a 
shallow, dispersing quasiparticle band and coherent/incoherent crossover at the HEA energy scale \cite{Zhang_anomaly,Fujimori_anomaly} for which 
solely extrinsic mechanisms are an inappropriate description \cite{Bansil_ME_effect}.

Here, new results from ARPES on Nd$_{1.83}$Ce$_{0.17}$CuO$_{4}$ (NCCO) also reveal the presence of a HEA.  These results demonstrate changes in the 
appearance of the band dispersion with incident photon energy (photoemission matrix elements) while clearly revealing an underlying band 
renormalization not attributable to matrix element effects.  The energy scale for the HEA, $\sim 500-750$ meV along the nodal direction, is 
approximately twice that found in the hole-doped high-$T_{c}$ compounds, as revealed in other studies 
\cite{CKim_anomaly,Fujimori_anomaly,Pan_anomaly}.  

To understand these results together with previous findings in hole-doped and parent cuprates, we present results from quantum Monte Carlo 
calculations of the single-band Hubbard model, across both sides of the phase diagram and at half-filling. Our results indicate that the HEA can be 
connected to doping and subsequent spectral weight transfers into the lower or upper Hubbard band of hole- or electron-doped cuprates, respectively.  
These calculations show that hole or electron doping away from the parent insulator promotes the formation of a quasi-particle band crossing $E_{F}$, 
the precursors of which can be found in the half-filled parent, and that the HEA represents a coherent/incoherent cross-over from this band to 
valence bands at higher binding energies.  In addition, the results highlight the electron/hole doping asymmetry of the HEA energy scale, in 
agreement with experimental observations. 

\section{Experimental Results}

\begin{figure}[th]
\centering
\includegraphics[width=4in]{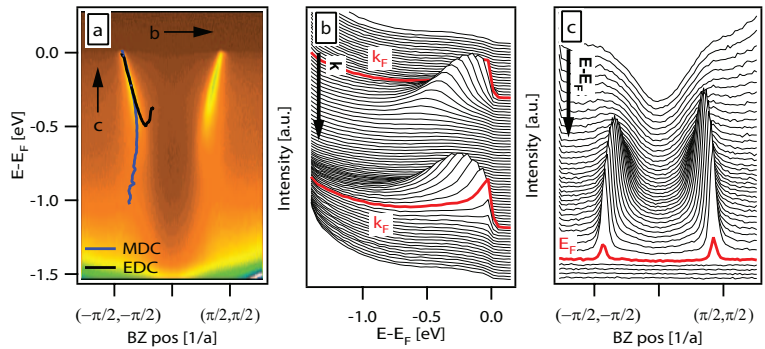}
\caption{\label{fig:HighEKinkNCCO.17}(a) Intensity plot along a nodal cut in the first BZ of NCCO (x=0.17) taken at $16.75$ eV incident energy.  (b) 
Energy distribution curves (EDCs) and (c) momentum distribution curves (MDCs) of the cut in (a).  The direction of the MDCs and EDCs are indicated by 
the black arrows in panel (a) and curves corresponding to $k_{F}$ and $E_{F}$ are highlighted in each panel.  The HEA appears as a ``waterfall" at an 
energy scale $\sim 500 - 600$ meV along this cut.}
\end{figure}

ARPES data were taken at beamline 5-4 of the Stanford Synchrotron Radiation Lightsource with a Scienta R4000 analyzer at photon energies from $15$ to 
$20$ eV. The energy resolution was $\sim 20$ meV with an angular resolution $\sim 0.3^\circ$. NCCO single crystals were grown in 4 atm of oxygen 
using the traveling-solvent floating-zone technique, annealed for 10 h in argon at 970 C followed by 20 h in oxygen at 500 C \cite{Motoyama} and then 
cleaved and measured at $10$ K and pressures better than $3\times10^{-11}$ torr. 

Figure 1 displays the ARPES data along a nodal cut ($(-\pi/a, -\pi/a)$ to $(\pi/a, \pi/a)$) for NCCO taken at $16.75$ eV.  The intensity plot of 
figure 1(a) clearly shows the HEA as a ``waterfall" in the dispersion at an energy $\sim 500 - 600$ meV.  This behavior is mirrored in the energy 
distribution curves (EDCs) and the momentum distribution curves (MDCs) shown in figures 1(b) and (c), respectively.  The EDC and MDC derived 
dispersions are superimposed over the intensity plot of figure 1(a).  This HEA demarcates the transition between a quasi-particle band at low binding 
energy and the oxygen valence bands at energies $\gtrsim 1$ eV.   While the MDCs appear to show a steep dispersion that bends back below the HEA, the 
EDCs show a shallower band with a 
rapid intensity decrease on the order of the HEA energy scale.  This difference is similar to that found in the half-filled parent insulators 
\cite{Ronning_anomaly} and hole-doped systems \cite{Graf_anomaly,Zhang_anomaly}.  

\begin{figure}[t!]
\centering
\includegraphics[width=4in]{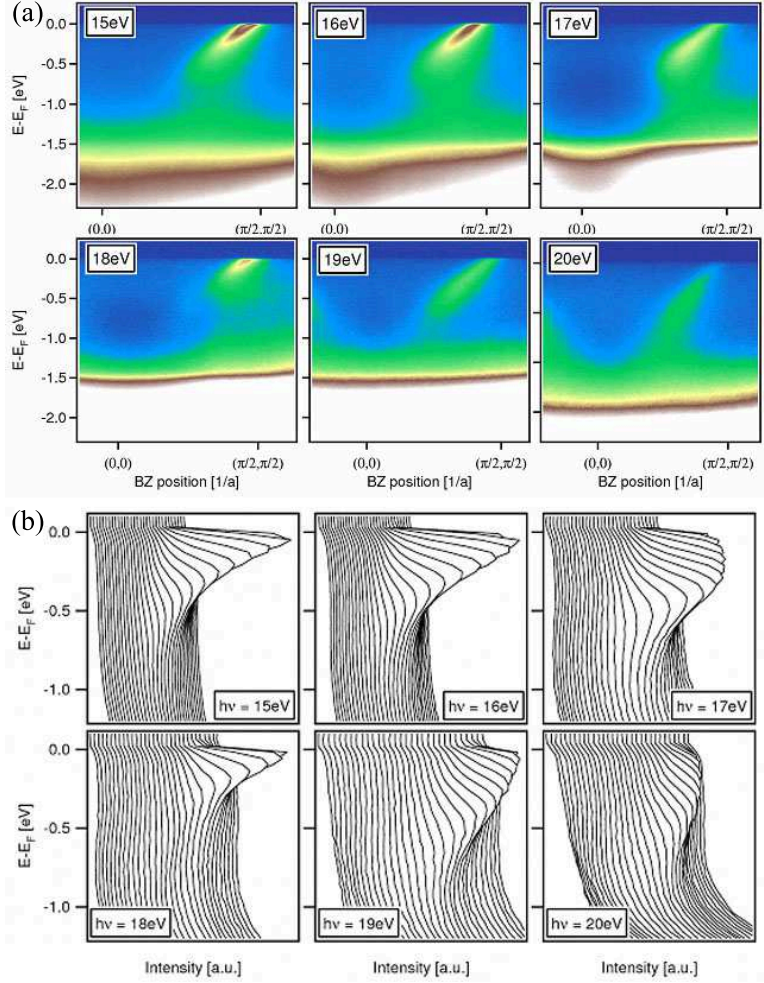}
\caption{\label{fig:HighEKinkNCCO.17_photon_energy}(a) Intensity plots along nodal cuts in the first BZ of NCCO (x=0.17) taken at the indicated 
incident photon energies.  Note the aesthetic changes induced by variation in matrix elements with photon energy.  (b) EDCs taken from the cuts in 
(a) at the same photon energies.  While the HEA appears as a ``waterfall" along these cuts for several photon energies, the EDC derived dispersions 
indicate a shallow, dispersing band with the HEA energy scale $\sim 500-750$ meV.}
\end{figure}

Figure 2(a) shows a series of nodal cuts at different photon energies highlighting the effect of changes to photoemission matrix elements.  
The data demonstrate 
aesthetic changes to 
features and apparent negative band velocities below the HEA energy scale 
intimately tied to matrix element effects. 
However, they also reveal an underlying band renormalization producing a HEA on the scale $\sim 500-750$ eV.  EDCs at these photon 
energies provide clarity to this point.  Figure 2(b) displays EDCs of the nodal cuts extracted from the respective data presented in 
figure 2(a).  Each 
of these EDCs shows a shallow dispersing band whose intensity decreases approaching the 
BZ center, similar to the behavior in hole-doped 
cuprates \cite{Graf_anomaly,Zhang_anomaly}.  EDC derived dispersions yield band bottoms 
(or the energy where the ARPES intensity falls-off)
in accord with the energy scale set by the HEA, not the deeper energy scales usually assigned by LDA estimates and MDC derived dispersions 
\cite{Graf_anomaly,Non_anomaly}.

\begin{figure}[t!]
\centering
\includegraphics[width=4in]{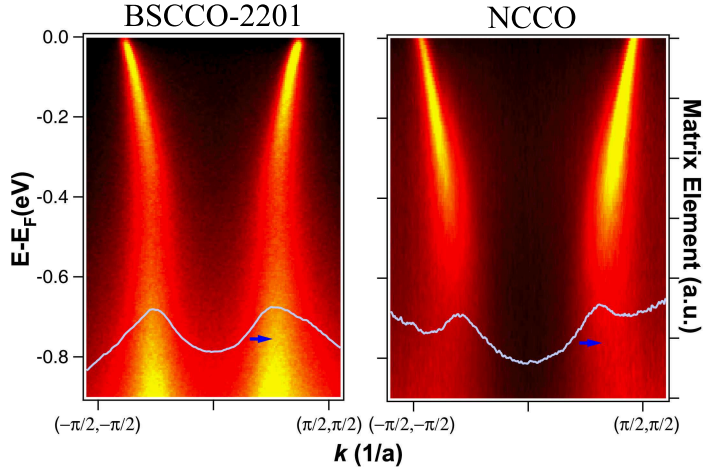}
\caption{Comparison of the HEA between hole- and electron-doped systems.  The left panel shows the nodal ARPES data for BSCCO-2201 adapted from 
\cite{Non_anomaly}; the right panel shows the nodal data for NCCO at $16.75$ eV from this study.  Matrix element profiles obtained from a 2D fitting 
procedure \cite{Non_fit} are also shown in each panel (the blue arrow indicates that the right most axis should be used for matrix element values).  While 
the appearance of the HEA is affected by changes in the photoelectric matrix elements, they can not account for the strong band renormalization.}
\end{figure}

A direct comparison of the ARPES data from Bi$_{2}$Sr$_{2}$CuO$_{6}$ (BSCCO-2201) \cite{Non_anomaly} and NCCO is shown in figure 3.  The HEAs are 
qualitatively similar in the two materials, but there are significant quantitative differences \cite{Non_anomaly,Pan_anomaly}.  Most importantly, 
this comparison highlights the fact that the HEA energy scale along the nodal direction is approximately twice that found in the hole-doped materials 
and the main valence band weight also lies at higher binding energy.

To address the issues surrounding matrix elements that vary with incident photon energy, a 2D fitting procedure, described in \cite{Non_fit}, was 
used to extract the matrix elements along the nodal cut in both hole- and electron-doped materials.  The momentum space profiles are shown in figure 
3.  In particular, the matrix elements significantly reduce the ARPES intensity upon approaching the $\Gamma$-point. However, taken together with the 
EDC and MDC dispersions for these materials, the matrix elements affect the appearance of the HEA, but can not account for the anomaly, also 
highlighted in a recent first-principles calculation \cite{Bansil_ME_effect}.

\section{Theoretical Results}

The experimental findings clearly indicate the presence of a HEA in NCCO, not attributable to matrix element effects.  This feature resembles the HEA 
found in both parent insulators \cite{Ronning_anomaly} and hole-doped systems 
\cite{Graf_anomaly,Non_anomaly,Feng_anomaly,Valla_anomaly,Chang_anomaly,Borisenko_anomaly,Zhang_anomaly}, and agrees well with the findings from 
previous studies on electron-doped compounds \cite{CKim_anomaly,Fujimori_anomaly,Pan_anomaly}.  This section provides a unifying description of the 
underlying mechanism that gives rise to the HEA within the simple framework of the single-band Hubbard model.

The two-dimensional, single-band Hubbard Hamiltonian 
serves as an effective, low energy model of the copper-oxide planes 
\cite{Zhang_Rice,Anderson} incorporating the effects of strong correlations.  The single-band Hubbard Hamiltonian has the form
\[
H=-\sum_{ij,\sigma}t_{ij}c^{\dag}_{i,\sigma}c_{j,\sigma}-\mu\sum_{i}n_{i}+\sum_{i}U(n_{i,\uparrow}-\frac{1}{2})(n_{i,\downarrow}-\frac{1}{2}),
\]
where $\{t_{ij}\}$ is a set of tight-binding parameters where only nearest-neighbor $t$ and next-nearest-neighbor $t'$ are different from zero, 
$c^{\dag}_{i,\sigma} (c_{i,\sigma})$ creates (annihilates) an electron with spin $\sigma$ at site $i$ and $n_{i,\sigma} = 
c^{\dag}_{i,\sigma}c_{i,\sigma}$ with $n_{i} = n_{i,\uparrow} + n_{i,\downarrow}$. The chemical potential $\mu$ controls the electron filling; and, 
in what follows, the Hubbard repulsion $U$ is set equal to the noninteracting electron bandwidth $W=8\,t$.   

Here, the single-band Hubbard model is studied using the determinant quantum Monte Carlo technique \cite{DQMC_1,DQMC_2}, an auxiliary-field method, 
to obtain the finite temperature, imaginary time propagator $G_{ij}(\tau)$ on a finite-size cluster with periodic boundary conditions.  The details 
of the method can be found in \cite{DQMC_1}.  The finite-size, square clusters used in this study have linear dimension $N=8$  and the imaginary time 
interval, partitioned into L ``slices" of size $\Delta\tau = \beta/L$, runs from 0 to $\beta = 1/T$, the inverse temperature. The hopping $t$ serves 
as the energy unit of the problem and unless otherwise noted $\Delta\tau = 1/16\,t\,$.  

The maximum entropy method \cite{Jarrell_Guber_MEM,Alex_MEM} provides an effective means of obtaining the spectral function $A(\mathbf{K},\omega)$ 
from the imaginary time data on the grid $\{\mathbf{K}\}$ in momentum space.  Once $A(\mathbf{K},\omega)$ has been obtained,  the single-particle 
self-energy $\Sigma(\mathbf{K},\omega)$ can be extracted using Dyson's equation and the tight-binding bandstructure.  Assuming a weak momentum 
dependence to the self-energy, an interpolation routine provides the value of $\Sigma(\mathbf{k},\omega)$ at an arbitrary point $\mathbf{k}$ in 
momentum space and Dyson's equation can be employed once again to compute $A(\mathbf{k},\omega)$. 

\begin{figure}[th]
\centering
\includegraphics[width=4in]{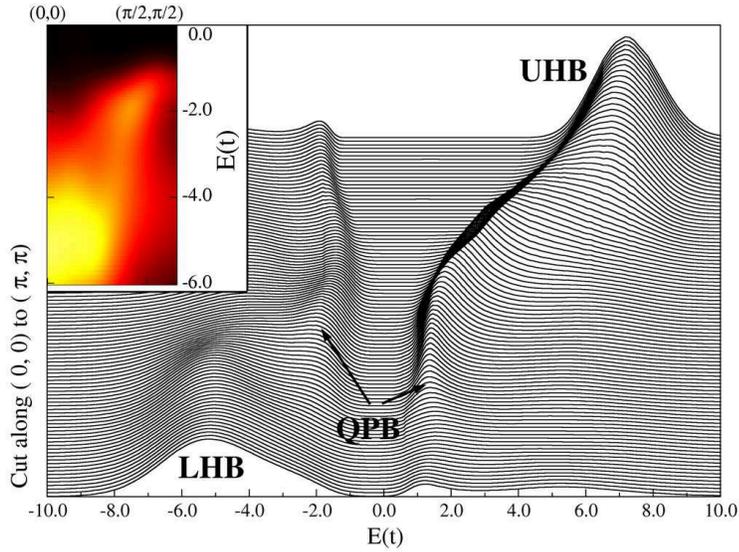}
\caption{Calculated single-particle spectral function $A(\mathbf{k},\omega)$ at half-filling ($\mu = 0.0\,t$), $t'=-0.3\,t$ and $\beta=3/t$, along a 
nodal cut from $(0,0)$ to $(\pi,\pi)$.  The inset shows a false-color plot of a portion of the same cut from $(0,0)$ to $\sim(\pi/2,\pi/2)$ below the 
Fermi level, qualitatively similar to experimental results on the parent insulators \cite{Ronning_anomaly}.  The labels denote the lower Hubbard band 
(LHB), upper Hubbard band (UHB), and quasi-particle-like branches (QPB) that serve as precursors to the bands that form, and cross $E_F$, with hole 
or electron doping.}
\end{figure}

The result from model calculations at half-filling provide a benchmark for the analysis and discussions of the behavior of the spectral function upon 
hole or electron doping.  Figure 4 shows $A(\mathbf{k},\omega)$ for the single-band Hubbard model at half-filling (parameters are given in the figure 
caption).  The canonical Hubbard bands are pronounced, but as in previous investigations \cite{Hanke_1,Hanke_2}, low energy satellite bands near 
$E_F$ also form.  These branches of the lower and upper Hubbard bands serve as precursors for the quasi-particle bands that form upon hole or 
electron doping.  These precursors also resemble the low energy features seen in ARPES experiments on the parent insulator \cite{Ronning_anomaly} 
which provides a link for comparing the HEA in hole- and electron-doped systems that follows.

\begin{figure}[th]
\centering
\includegraphics[width=4in]{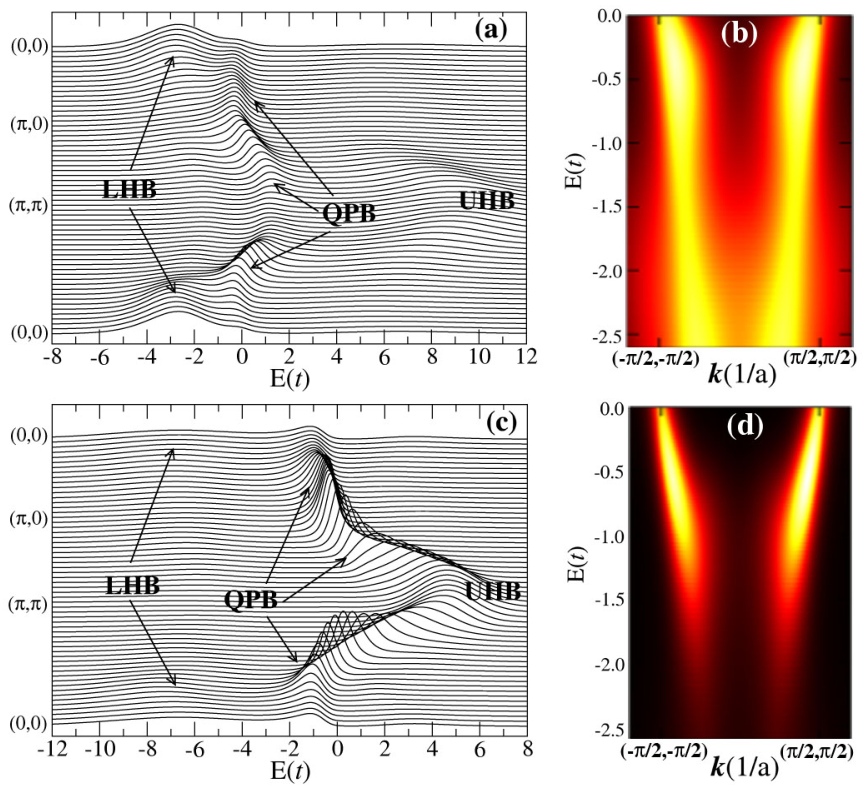}
\caption{Calculated single-particle spectral function $A(\mathbf{k},\omega)$ for $\sim 14\%$ hole-doping ($\mu = -2.5\,t$), $t'=-0.3\,t$ and 
$\beta=3/t$ in panels (a) and (b) and $\sim 16\%$ electron-doping ($\mu = 2.25\,t$), $t'=-0.2\,t$ and $\beta=3/t$ in panels (c) and (d).  Panels (a) 
and (c) trace out the path $(0,0)-(\pi,\pi)-(\pi,0)-(0,0)$ in the first BZ.  Panels (b) and (d) focus on nodal cuts from $\sim(-\pi/2,-\pi/2)$ to 
$\sim(\pi/2,\pi/2)$ where representative photoelectric matrix elements have been used in creating the intensity plots.  The labels in panels (a) and 
(c) refer to the lower Hubbard band (LHB), upper Hubbard band (UHB), and quasi-particle band (QPB) with arrows provided to clarify the assignments.}
\end{figure}

Results from calculation of $A(\mathbf{k},\omega)$ for the single-band Hubbard model upon doping are shown in figures 5(a)-(d).  The single-particle 
spectral function is shown for a representative system with parameters $t'=-0.3\,t$ and $\beta = 3/t$ at $\sim 14\%$ hole doping ($\mu = -2.5\,t$) in 
figures 5(a) and (b) while (c) and (d) show the results for a system with parameters $t'=-0.2\,t$ and $\beta = 3/t$ at $\sim 16\%$ electron doping 
($\mu = 2.25\,t$).  For each system the tight-binding parameters and even Fermi surface topology are quite similar.  However, note the two sets of 
parameters are not simply related by a particle-hole transformation and, in general, the experimentally determined bandstructure parameters would not 
reveal such a symmetry between hole- and electron-doped materials.

For the hole-doped system the lower Hubbard band (LHB) is essentially localized near the $\Gamma$-point with a weak tail of spectral intensity 
extending toward the points $(\pi,\pi)$ and $(\pi,0)$.  The decrease in spectral weight in these tails approximately coincides with the location in 
momentum space identified with the HEA.  The weak appearance of the upper Hubbard band (UHB) results from hole-doping 
\cite{Eskes_LDA+U,Sawatzky_LDA+U} as spectral weight is transferred into the LHB or more precisely the dispersive quasi-particle like band (QPB) that 
develops from low binding energy precursors found at half-filling \cite{Hanke_1,Hanke_2} seen in figure 4.

Along a nodal cut, the QPB is highly dispersive crossing the Fermi level at $\sim (\pi/2,\pi/2)$.  Near $(\pi/4,\pi/4)$ the spectral intensity in the 
band drops and the HEA appears as an apparent cross-over from the QPB to the LHB at an energy $\sim -0.5\,t$ to $-0.75\,t$, the HEA energy scale.  
While the spectral intensity decreases as the QPB approaches the $\Gamma$-point, the coexistence of the LHB and QPB over a significant momentum 
interval supports the cross-over scenario presented here. 

In the electron-doped system, the dispersive QPB shown in figure 5(c) dips farther below the Fermi level than its hole-doped counterpart, to an 
energy $\sim -1.0\,t$ to $-1.5\,t$, in agreement with the comparison shown in figure 3. More importantly, the QPB arises from intermediate energy 
precursors in the UHB of the half-filled system \cite{Hanke_1,Hanke_2}, in contrast to the spectral weight forming the hole-doped QPB.  While there 
appears to be a weak tail of the UHB at significantly higher energies ($\sim 3\,t$ to $4\,t$), the dispersive feature and bulk of the UHB appear to 
meet in another ``waterfall" well above the Fermi level.

The appearance of the HEA as a ``waterfall" at intermediate binding energies may be highlighted by applying photoelectric matrix elements, 
representative of experimentally derived values, to the results obtained from the calculations.  Figures 5(b) and (d) show $A(\mathbf{k},\omega)$ 
multiplied by these matrix elements (see \cite{Non_fit}) along a nodal cut for hole- and electron-doped systems, respectively.  In each case the 
matrix elements severely suppress the intensity near the $\Gamma$-point.  In particular for hole-doped systems, this behavior is consistent with the 
interpretation of the QPB as a renormalized Zhang-Rice singlet (ZRS) band with photoelectric matrix elements that also would reflect the robustness 
of the Zhang-Rice picture in different regions of the BZ \cite{Zhang_Rice,Savrasov_ZRS}.  This suppression of intensity, particularly in the 
hole-doped calculation, leads to the appearance of the HEA as a ``waterfall" due to the overlap of spectral weight at intermediate energies for 
$\mathbf{k}$ away from the $\Gamma$-point.  In the electron-doped calculation, the lack of significant spectral weight in the LHB leads to a simple 
suppression of intensity near the $\Gamma$-point without a significant overlap at intermediate energies.  This result should be affected by the 
inclusion of additional valence band weight within multi-band treatments.

\section{Discussion}

\begin{figure}[th]
\centering
\includegraphics[width=4in]{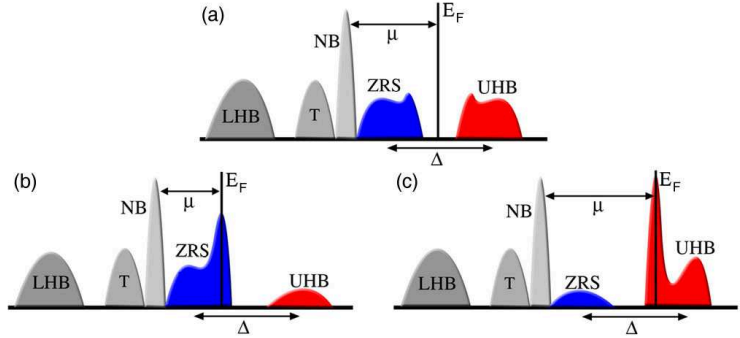}
\caption{Schematic diagram for the multi-band structure of the copper-oxide plane highlighting the Zhang-Rice singlet (ZRS) band (blue) and upper 
Hubbard band (UHB) (red).  The schematic also includes the multi-band lower Hubbard band (LHB), the Zhang-Rice triplet band (T) and the oxygen 
non-bonding band (NB).  The ZRS band serves as the LHB in the single-band model. (a) Bands at half-filling highlighting the incoherent portions of 
the ZRS band (LHB for single-band) and UHB together with the quasi-particle band precursors.  The chemical potential $\mu$ (Fermi level $E_{F}$) 
shifts as a function of (b) hole or (c) electron doping into the ZRS band (LHB for single-band) or the UHB, respectively.}
\end{figure}

The relative spectral weight in the LHB and UHB as well as the HEA energy scale exemplify the dichotomy between hole- and electron-doped systems.  A 
schematic of the band structure can be found in figure 6. In a multi-band treatment incorporating planar oxygen degrees of freedom, the UHB and ZRS 
band correspond roughly to the UHB and LHB in the single-band Hubbard model.  Additional bands (proper multi-band LHB, Zhang-Rice triplet and non-bonding 
oxygen) serve as marker states and play no role in the single-band calculations.  

Upon hole doping in the single-band model, the chemical potential shifts into the LHB with a concomitant transfer of spectral weight from the UHB and 
increase in weight at energies near the chemical potential, forming the QPB (see figure 6(b) and changes to the multi-band ZRS band and UHB).  In 
contrast, upon electron doping, the chemical potential shifts into the UHB with a transfer of weight to higher relative energy, again forming the 
bulk of the QPB (see figure 6(c)).  The chemical potential and spectral weight shifts with doping are consistent with the treatment presented in 
Ref.~\cite{Eskes_LDA+U}. Note with doping that the Mott gap does not simply collapse, but instead forms much of the intermediate energy regime below 
the HEA energy scale in the electron-doped calculation.  This is in contrast to the hole-doped system in which the anomaly lies within the canonical 
LHB and the Mott gap has been effectively pushed above the Fermi level.

The presented results support the conclusion that strong correlations and many body effects, here in the guise of the single-band Hubbard model, play 
a central role in the high energy anomaly.  The calculations also echo some of the results of much earlier investigations into the spectral function 
of the single-band Hubbard model \cite{Hanke_1,Hanke_2,Dagotto_RMP}.  This study shows evidence for an anomaly in electron-doped materials and does 
not equate its energy scale, under either hole or electron doping, in any simple way with $J$ that should be approximately equivalent in the two 
types of materials.  
Band renormalization due to strong correlations incorporates coupling to paramagnon modes at these high energy scales, combined with all other 
renormalization pathways, to produce 
a transition, or cross-over, from a quasi-particle band at low binding energy near $E_{F}$ to the incoherent lower Hubbard band, or more precisely 
oxygen valence bands, at higher binding energy. This behavior is consistent with experimental observations and not captured in weak coupling 
approaches that produce a simple ``kink" in the dispersion at the HEA energy scale \cite{QMC,Paramagnon_anomaly,Bansil_ME_effect}.  While similar 
conclusions have been reached in Ref.~\cite{Fujimori_anomaly}, the combined experimental and theoretical treatment presented here elucidates the 
complex interplay between doping, spectral weight transfer, and band renormalization that gives rise to the HEA beyond a straightforward chemical 
potential shift and a tight-binding analysis.

Qualitatively, the doping and momentum dependence obtained from calculations are similar to the experimental results and, in many cases, minor 
adjustments to model parameters produce even surprisingly close quantitative agreement with the observed features 
\cite{Graf_anomaly,Non_anomaly,Feng_anomaly,Valla_anomaly,Chang_anomaly,Borisenko_anomaly,Zhang_anomaly,CKim_anomaly,Fujimori_anomaly,Pan_anomaly}, 
especially those presented here on electron-doped NCCO.  
In the near nodal region, the agreement depends far less on the fine details of the underlying bandstructure; while moving toward the anti-nodal 
region, a quantitative agreement between theory and experiment depends heavily on the tight-binding fit parameters due to the already shallow band
dispersion and relative strength of the renormalization, also demonstrated in the analysis of Ref.~\cite{Fujimori_anomaly}.  While a great deal of 
information has been obtained about the hole doping dependence of the HEA \cite{Graf_anomaly,Non_anomaly}, the electron doping dependence remains largely
unexplored to this point as do effects associated with changing matrix elements between different BZs.  Preliminary results indicate that our theoretical
simulations are qualitatively robust to changes in electron doping, as they are to hole doping, and also qualitatively reproduce dispersions in other
BZs, but this remains the subject of ongoing investigation and detailed results will appear in a separate publication.  It is also important to note that
the single-band Hubbard model should only be viewed as a low energy effective model of the cuprates.  A better understanding of the role played by 
strong correlations would come from careful analysis of calculations explicitly incorporating separate copper and
oxygen 
degrees of freedom that would capture the formation of Zhang-Rice singlets \cite{Zhang_Rice,Savrasov_ZRS} and the spectral weight associated with 
the valence bands, assumed to have strong oxygen character.

\ack

The authors would like to thank A. Fujimori, R.~Hackl, M.~Jarrell, W.~S.~Lee, A.~Macridin, T.~Maier, G. Sawatzky, and F.~Vernay for valuable 
discussions. This work was supported by the U.S. Department of Energy, Office of Basic Energy Sciences under contracts DE-AC02-76SF00515 and 
DE-FG03-01ER45929-A001 and Office of Advanced Scientific Computing Research through the Scientific Discovery through Advanced Computing (SciDAC) 
program under contract DE-FC02-06ER25793, along with support from the National Science Foundation under contract DMR-0705086 and the Natural Sciences 
and Engineering Research Council of Canada (NSERC).  CK acknowledges support from KICOS in No. K20602000008.  Portions of this research were carried 
out at the Stanford Synchrotron Radiation Lightsource (SSRL), a national user facility operated by Stanford University on behalf of the U.S. 
Department of Energy, Office of Basic Energy Sciences.  The computational work was made possible by resources of the National Energy Research 
Scientific Computing Center (NERSC), which is supported by the Office of Science of the U.S. Department of Energy under Contract No. 
DE-AC02-05CH11231, the National Science Foundation through TeraGrid resources provided by the National Center for Supercomputing Applications (NCSA), 
R\'{e}seau Qu\'{e}b\'{e}cois de Calcul de Haute Performance (RQCHP) through Mammouth-parall\`{e}le at the University of Sherbrooke and the facilities 
of the Shared Hierarchical Academic Research Computing Network (SHARCNET). Three of the authors (BM, SJ, TPD) wish to thank the Walther Mei{\ss}ner 
Institute (WMI) and the Pacific Institute of Theoretical Physics (PiTP) for their hospitality during part of this work.

\section*{References}

\bibliographystyle{iopart-num}
\bibliography{Moritz_HEA}

\end{document}